\documentclass[showpacs
,showkeys
,nofootinbib
,amsfonts
,floatfix
,twocolumn
,tightenlines
,preprintnumbers
,aps
,prd
,longbibliography
,superscriptaddress
]{revtex4-2}
\pdfoutput=1

\usepackage{adjustbox} 
\usepackage{graphicx}  
\usepackage{mathtools}
\usepackage{amsmath, amssymb, bm,bbm}   
\usepackage{bbold}
\usepackage{xcolor}
\usepackage{natbib}
\usepackage{hyperref}
\usepackage{csvsimple} 
\usepackage{ifthen}
\usepackage{siunitx} 
\usepackage{tikz,adjustbox}

\usetikzlibrary{quantikz2}
\usepackage[T1]{fontenc}
\usepackage{physics}

\newcommand{\bn}{\mathbf n}
\newcommand{\eq}[1]{Eq.~(\ref{#1})}
\newcommand{\fig}[1]{Fig.~\ref{#1}}
\newcommand{\tab}[1]{Table~\ref{#1}}

\newcommand{\dblquotes}[1]{``#1''} 

\begin{document}
\preprint{FERMILAB-PUB-25-0849-T}
\title{Quantum computation of mass gap in an asymptotically free theory}

\author{%
Paulo F. Bedaque}
\email{bedaque@umd.edu}
\affiliation{Department of Physics, University of Maryland, 
College Park, MD 20742, USA}

\author{%
Edison M. Murairi}
\email{emurairi@fnal.gov}
\affiliation{Superconducting and Quantum Materials System Center (SQMS), Batavia, Illinois, 60510, USA.}
\affiliation{Fermi National Accelerator Laboratory, Batavia, Illinois, 60510, USA}

\author{%
Gautam Rupak}
\email{grupak@ccs.msstate.edu}
\affiliation{Department of Physics \& Astronomy and HPC$^2$ Center for 
Computational Sciences, Mississippi State
University, Mississippi State, MS 39762, USA}

\author{%
Valery S. Simonyan}
\email{vsimon24@umd.edu}
\affiliation{Department of Physics, University of Maryland, 
College Park, MD 20742, USA}

\begin{abstract}
In relativistic field theories, the mass spectrum is given by the difference between the energy of the vacuum and the excited states. Near the continuum limit, the cancellation between these two values leads to  loss of precision. We propose a method to extract the mass gap directly using quantum computers and apply it to a
particular version of the nonlinear $\sigma$-model with the correct continuum limit and perform calculations in quantum hardware (at strong coupling) and simulation in classical computers (at weak coupling).

\end{abstract}

\maketitle

\section{Introduction}
\label{sec:Introduction}

A central challenge in computational Quantum Chromodynamics (QCD), and in numerical quantum field theory in general,  is the first-principles calculation of physically important but inaccessible observables within the standard lattice formulation. The Euclidean path integral approach, while highly successful for a wide range of zero-density and static quantities, encounters a fundamental obstruction known as the sign problem when a finite baryon chemical potential is introduced to study dense matter. The ensuing complex action prevents efficient Monte Carlo sampling, precluding ab initio exploration of the QCD phase diagram at high density. A related, and equally profound, challenge is the calculation of real-time observables. Key quantities such as transport coefficients, scattering amplitudes, and parton distribution functions are defined in Minkowski spacetime and cannot be directly extracted from imaginary-time lattice correlators without an ill-posed analytic continuation.

Quantum computations of QCD observables would in principle avoid the aforementioned problems by directly performing the calculations in Minkowski space using unitary time evolution with the Hamiltonian. Formulating QCD and other quantum field theories in a form amenable to quantum computation, as well as developing practical algorithms for their simulation, is a very active field of research. In one approach, the bosonic fields are truncated in the Fourier basis~\cite{PhysRevA.73.022328,Kan:2022esj,Kan2021,Balaji2025,Zohar2013,Zohar2017,Zohar2015,Ciavarella2025} or a mixed basis~\cite{Grabowska2025}. In another approach, the underlying continuous symmetry group is replaced with a discrete subgroup~\cite{Lamm2019,Lamm2024,Gustafson2021,Gustafson2022,Gustafson2024,Alam2022,Murairi2024,Gustafson2024a,Perez:2025cxl}. Other methods include a fuzzy regulation~\cite{Alexandru:2022son,Alexandru:2021xkf,fuzzy-gauge}, quantum link models~\cite{PhysRevD.106.L091502,CHANDRASEKHARAN1997455,Wiese:2021djl}.
A common framework for these calculations involves a multi-stage process: the first is encoding the theory's degrees of freedom into a quantum register; second, preparing the ground state (or a close approximation of it); third, implementing time evolution generated by the Hamiltonian in terms of quantum gates; and finally, measuring the relevant observables.
Significant theoretical challenges are inherent to each of these steps, presenting obstacles above and beyond the challenging engineering problem of building large, fault-tolerant quantum computers.

In this work, we address the challenge of computing a fundamental property of strongly-coupled field theories—the mass gap—using a quantum computer. As a concrete testbed for our method, we employ the (1+1)-dimensional nonlinear $O(3)$ $\sigma$-model. This theory provides an ideal proving ground, as it shares key non-perturbative features with QCD, namely asymptotic freedom and dynamical mass generation, while being more amenable to initial simulations on current quantum hardware.
For that, some issues have to be addressed. The first, already tackled in \cite{Alexandru:2019ozf,Alexandru:2021xkf,Alexandru:2022son,shailesh-1,Bhattacharya:2020gpm,shailesh-3}, arises from the fact that this theory contains bosonic fields, with an infinite number of degrees of freedom at each space-time point, which  cannot be accommodated in a finite quantum register. 
This problem was solved by finding a theory that can be encoded in a finite number of qubits per lattice site (only 2 in fact) which, neverthless, belongs to the same universality class as the full bosonic $O(3)$ nonlinear $\sigma$-model. This theory, referred to as the \dblquotes{fuzzy $\sigma$-model} \cite{Alexandru:2019ozf,Alexandru:2021xkf,Alexandru:2022son} or the \dblquotes{Heisenberg comb} \cite{shailesh-1,Bhattacharya:2020gpm,shailesh-3} is what we simulated in a quantum computer.
The second issue we address comes from the fact that the mass gap (the mass of the lightest particle) in lattice units approaches zero in the continuum limit. Computing this number accurately from the difference between the {\it approximate} values of the much larger ground  and the first excited state energies $E_0, E_1$ is a difficult feat leading to a severe loss of precision. We suggest, implement, and test a way around this problem.

\section{The fuzzy \texorpdfstring{$\bm{O(3)}$}{O3}  $\sigma$-model }
\label{sec:O3}

For applications in quantum computing, field theories are most naturally formulated in the Hamiltonian framework.
 Thus, we will start briefly reviewing the (regular, non-\dblquotes{fuzzy}) $1+1$ dimensional $O(3)$  nonlinear $\sigma$-model in that language.
Consider a one-dimensional spatial lattice of $L$ sites separated by a lattice spacing $ a $. At each site $x$, the local Hilbert space of the model consists of square-integrable functions over the unit sphere $S^2$, $\psi(\bn)$, $\bn^2=1$. These functions can be expanded in an infinite series of $\bn$ as
\begin{equation}\label{eq:taylor_sigma}
\psi(\bn) = \psi_0 + \psi_i \bn_i + 
\psi_{ij} \bn_i \bn_j+ \cdots\,,
\end{equation}
where $\bn_i$ are the components of $\bn$. The full Hilbert space is the tensor product over all sites.
The field  operators $\mathbbm{n}_i(x)$ act by multiplication: $\mathbbm{n}_i(x)\psi = \bn_i(x)\psi$. Global $O(3)$ rotations act as $\mathbb{R}\psi(\{\bn(x)\}) = \psi(\{R^{-1}\bn(x)\})$ and are generated by angular momentum operators $\mathbb{T}_k$. The on-site algebra satisfies
\begin{align}\label{eq:comm-sigma}
[\mathbbm{n}_i, \mathbbm{n}_j] &= 0\,,\nonumber \\
[\mathbb{T}_i, \mathbb{T}_j] &= i \epsilon_{ijk} \mathbb{T}_k\,,\nonumber \\
[\mathbb{T}_i, \mathbbm{n}_j] &= i \epsilon_{ijk} \mathbbm{n}_k\,.
\end{align}
The unit norm constraint is enforced as restriction on the Hilbert space:
\begin{equation}
\left[\mathbbm{n}(x)^2 - \openone\right]\psi = 0\, ,
\end{equation}
and the Hamiltonian is given by \cite{Bruckmann:2019} \footnote{We use units where $\hbar=1=c$.}:
\begin{equation}\label{eq:H_sigma}
 a  \mathbb{H} = \eta g^2 \sum_x \mathbb{T}(x)^2 - \frac{\eta}{g^2} \sum_x \mathbbm{n}(x) \cdot \mathbbm{n}(x+1)\,,
\end{equation} is a sum of a \dblquotes{kinetic} term and a nearest-neighbor \dblquotes{potential} energy. Here $a$ is the spatial lattice spacing and $\eta$ and $g^2$ two parameters controlling the strength of each term. 

The coefficients in \eq{eq:H_sigma} were parametrized in order to make its eigenstates (and, consequently, the correlation length $\xi$)  independent of $\eta$, while energies are simply proportional to it. Lorentz invariance is restored by tuning $\eta$ so the energy gap $E_1-E_0$ equals the mass of the lightest particle $m=1/\xi$. The continuum limit ($\xi \gg  a $) is obtained by setting $g( a )\to 0$ as $ a \to 0$
according to
\begin{equation}
    \frac{dg( a )}{d\log  a } = b_0 g^2\qty(a)\,,
\end{equation} where $b_0$ is a known numerical constant \cite{polyakov}. Since the energy gap is independent of $ a $ in the continuum limit, it must have the form
\begin{equation}\label{eq:gap_pert}
     m \sim \frac{1}{a}
     \exp\left[-\frac{1} {2b_0g^2\qty(a) }\right]\,.
\end{equation}
In the continuum limit the nonlinear model is solvable and it describes a triplet of particles of mass interacting through a $\delta$-function interaction \cite{zamolodikov}. Usually, one sets the value of $g(a)$ at a particular value of $a=\bar a$ by demanding that the experimentally known particle mass $m$ is reproduced. After fixing $g(\bar a)$, all other observables can be computed, a procedure known as \dblquotes{scale setting}. 

The Hilbert space of this theory is infinite dimensional even if $L$ is finite. Therefore, its states cannot be encoded in a finite quantum register, a feature shared by all bosonic field theories. Therefore, it is necessary to truncate not only space (that is what the lattice regularization does), but also {\it field} space (in this case the unit sphere), a process sometimes called \dblquotes{qubitization}.
Various qubitizations of the $\sigma$-model have been proposed. One is to substitute the unit sphere $S^2$ by the vertices of one of the Platonic solids, in order to maintain as much of the rotation symmetry as possible. The resulting model, however, does not have a continuum limit \cite{Caracciolo_2001,Patrascioiu_1998,Hasenfratz:2001iz}. Another possibility is to expand the wavefunction at each site  into spherical harmonics but truncate the series to $\ell \leq \ell_{\text{max}}$ \cite{Bruckmann:2019}. Despite keeping the full rotation invariance, the resulting model also lacks a continuum limit for fixed $\ell_{\text{max}}$, as shown in \cite{Alexandru:2022son}.
But a qubitization with the proper continuum limit exists. It was originally \cite{Alexandru:2019ozf} inspired by a construction in non-commutative geometry, the \dblquotes{fuzzy sphere}~\cite{hoppe2002membranes,Madore:1991bw} and, at the same time, by very different considerations \cite{Singh:2019uwdsuu}. Further studies using Monte Carlo and tensor network methods \cite{Alexandru:2021xkf,Bhattacharya:2020gpm,Liu:2021} established that the continuum limit of the fuzzy model  is asymptotically free and indeed coincides with regular $\sigma$-model defined by \eq{eq:H_sigma}. The fuzzy $\sigma$-model is defined by choosing an irreducible representation $j$ of the symmetry group $SU(2)$. For simplicity here, we will choose $j=1/2$. The Hilbert space of the regular $\sigma$-model is substituted by the (finite dimensional) space
of $2\times 2$ matrices.\footnote{One can also use the generators of larger irreps. and a correspondingly larger Hilbert space.} The  analogue of \eq{eq:taylor_sigma} is the finite expansion 
$\Psi = \psi_0 \openone + \psi_k \sigma_k$, where $\sigma_k=(X,Y,Z)$ are the Pauli matrices, is the finite-dimensional. 

The operators $\mathfrak{n}_k = \frac{1}{\sqrt{3}} \sigma_k$ and $\mathcal{T}_k =\frac{1}{2} [\sigma_k, \cdots]$ are the fuzzy analogues of $\mathbbm{n}_k, \mathbb{T}_k$ and satisfy
\begin{align}
    \mathfrak{n}^2 &= \mathbb{1}\,,  
   & [\mathfrak{n}_i, \mathfrak{n}_j ] &= \frac{2}{\sqrt{3}}\epsilon_{ijk} \mathfrak{n}_k\,,\nonumber\\
[\mathcal{T}_i, \mathcal{T}_j ] &= i\epsilon_{ijk} \mathcal{T}_k\,,
&[\mathcal{T}_i, \mathfrak{n}_j ]& = i\epsilon_{ijk} \mathfrak{n}_k\,.
\end{align} Except for the commutation relation among the $\mathfrak{n}_i$, those are the same as in  \eq{eq:comm-sigma}. This is the sense in which the unit sphere of the $\sigma$-model was substituted by a non-commutative manifold \footnote{Note that it is the field space that is non-commutative, not the spacetime as in a completely different class of theories extensively studied \cite{nekrasov}.}.
The Hamiltonian of the fuzzy model is the analogue of \eq{eq:H_sigma}:
\begin{equation}
 a  H = \eta g^2 \sum_x \mathcal{T}(x)^2 + \frac{\eta}{g^2} \sum_x \mathfrak{n}(x) \cdot \mathfrak{n}(x+1)\,.
\label{eq:H_fuzzy}
\end{equation} Notice the sign change between \eq{eq:H_sigma} and \eq{eq:H_fuzzy}: the nearest-neighbor interaction in \eq{eq:H_fuzzy} is 
antiferromagnetic. It is convenient to expand the wavefunction at every site as 
$\Psi = \mathcal{\psi}_\mu E_\mu$ with 
\begin{equation}
    e_{0} \!= \!\begin{bmatrix}
        1 & 0\\
        0 & 0
    \end{bmatrix},  
    e_{1}\! = \!\begin{bmatrix}
        0 & 1\\
        0 & 0
    \end{bmatrix}, 
    e_{2}\! =\! \begin{bmatrix}
        0 & 0\\
        1 & 0
    \end{bmatrix}, 
    e_{3} \!=\! \begin{bmatrix}
        0 & 0\\
        0 & 1
    \end{bmatrix}.
\end{equation}

\begin{figure}[tbh]
\begin{center}
\includegraphics[width=0.45\textwidth,clip=true]{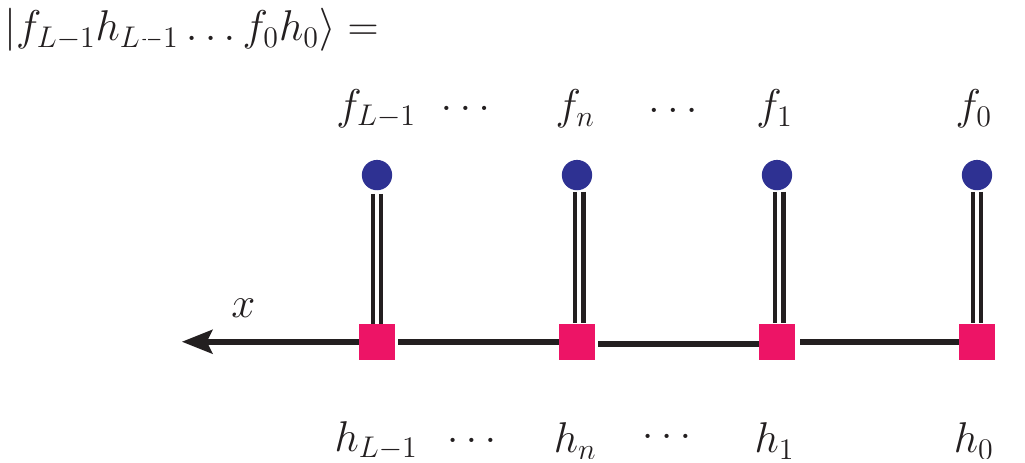}
\end{center}
\caption{\protect Diagram depicting the \dblquotes{Heisenberg comb} form of the fuzzy $\sigma$-model~\cite{Bhattacharya:2020gpm}. The \dblquotes{head} qubits are denote by $h_n$ (red squares) and the \dblquotes{fuzz} qubits are denoted by $f_n$ (blue circles) at site $l=0$, $1,\dots$, $L-1$.}
\label{fig:FuzzySigma}
\end{figure}

After a further conjugation with $\openone\otimes\sigma_2$ at each site we arrive at the following form for the Hamiltonian (this is in the form used in \cite{Bhattacharya:2020gpm}):
\begin{multline}\label{eq:fuzzyNew}
     a  H = \frac{\eta g^2}{2} \sum_{l=0}^{L-1}
    [X_{2l+1} X_{2l}
    +
    Y_{2l+1} Y_{2l}+Z_{2l+1} Z_{2l}]\\
    + \frac{\eta}{3 g^2}\sum_{l=0}^{L-1}[X_{2l+2}X_{2l}+Y_{2l+2}Y_{2l}+Z_{2l+2}Z_{2l}]\,,
\end{multline} 
where periodic boundary condition in the second summation of terms is implied. This Hamiltonian differs from the one in \eq{eq:H_fuzzy} only by a term proportional to the identity that just shifts the energy eigen values without affecting the gap or the eigen vectors.
    Equation \eq{eq:fuzzyNew} can be visualized using the schematic in \fig{fig:FuzzySigma} \cite{Bhattacharya:2020gpm}, where the even and odd qubits—labeled the \dblquotes{head} (squares) and \dblquotes{fuzz} (circles) respectively—are coupled via single and double lines representing the anti-ferromagnetic couplings in the potential and kinetic terms, respectively.

    The model defined by \eq{eq:fuzzyNew} is easily understood in two limits. First, in the strong coupling limit $g\rightarrow\infty$, only the kinetic term survives and the ground state factorizes into a tensor product of singlets at each site:
\begin{equation}\label{eq:strong_ground}
    \ket{\Psi_\text{strong}} = \bigotimes_{l=0}^{L-1} 
    \frac{\ket{0_{2l+1}1_{2l}}-\ket{1_{2l+1}0_{2l}}}{\sqrt{2}}\,,
\end{equation} where $\ket{q_l}$ with $q_l=0,1$ correspond to the state of the $l$-th qubit. Conversely, in the weak coupling limit $g\rightarrow 0$,  the potential term dominates.  In this limit, the fuzz qubits completely decouple and   the ground state has a $2^L$-fold degeneracy corresponding to all possible states of the fuzz qubits. In fact,   all states of the form $\ket{\Psi} = \ket{\text{antiferro}} \otimes \ket{\text{any}}$,
 where $\ket{\text{antiferro}}$ is the ground state of the Heisenberg antiferromagnetic chain among the head qubits and $\ket{\text{any}}$ being any state of the fuzz qubits, are degenerate in the weak coupling limit. 
 The kinetic term lifts this degeneracy only at second order in perturbation theory \cite{fuzzy-gauge}. We find that, of all states of this form, the state that smoothly connects to the ground state at non-zero $g$ is
\begin{equation}\label{eq:weak_ground}
    \ket{\Psi_\text{weak}} = \ket{\text{antiferro}} \otimes \ket{\text{antiferro}}\,.
\end{equation} As explained in Ref.~\cite{fuzzy-gauge}, perturbing the ground state $\ket{\psi_\text{weak}}$ in \eq{eq:weak_ground} with the kinetic term to obtain the weak coupling expansion series leads to infrared divergences as $L\rightarrow\infty$, a hallmark of asymptotically free theories.

\section{Measuring the gap}
\label{sec:algorithm}

 In all relativistic field theories, the energy of the ground state has no physical meaning. The interest is instead on observables like the mass of the lightest particle, given by the energy gap between the ground and the first excited state. This poses a challenge to quantum computation. 
Near the continuum limit, the ground state energy $E_0$ and the excited state energy (or degenerate energies) $E_1$ are of the order of the inverse lattice spacing $\sim 1/a$ while their difference $\omega=E_1-E_0$ is much smaller $\omega \sim a^0$. Any {\it approximate} determination of the individual state energies leads to large deviations in the mass gap.
 In this work, we introduce a paradigm that avoids this pitfall by constructing an observable for measuring the mass gap directly.
 Suppose  we setup a linear combination of the ground $\ket{E_0}$ and first excited $\ket{E_1}$ states, respectively,   as an initial condition:
\begin{equation}
    \ket{\Psi(0)} = \alpha \ket{E_0} + i \beta \ket{E_1}\,.
\end{equation} 
At later times, the expectation value of the operator $d=\ket{E_0}\bra{E_1}+\ket{E_1}\bra{E_0}$ connecting these two states is given by 
\begin{equation}\label{eq:sin}
    \bra{\Psi(t)} d \ket{\Psi(t)}
    =
    -2\alpha\beta \sin\left(m t \right)\,,
\end{equation}
where $\alpha$, $\beta$ are assumed to be real.   Therefore, measuring $\langle d \rangle$ as a function of time gives the value of the mass gap directly, bypassing the delicate cancellation between $E_1$ and $E_0$. Of course, knowing the states $\ket{E_0}, \ket{E_1}$ amounts to solving the theory. The practical strategy is to have a rough approximation of these states so the oscillations in $\langle d \rangle$   are dominated by \eq{eq:sin}. Other, higher frequency oscillations, can be filtered out by a number of techniques as long as their amplitude is not too large, as we will demonstrate below. We refer to $d$ as the dipole operator, as it is reminiscent of an electric dipole transition operator connecting states with opposite parity, in this case singlet and triplet states.

Note that the expectation value of any operator that does not commute with the Hamiltonian $H$ would show oscillations corresponding to the various relative phases associated with energy differences in a superposition state of energy eigenstates. To see a clear signal of the oscillation due to $m=E_1-E_0$, an initial state composed mostly of $\ket{E_0}, \ket{E_1}$ states, and an operator mostly coupling these states is preferable. 

In the strong coupling region $g\agt 1$ we approximate the ground state by its strong coupling limit (\eq{eq:strong_ground}):
\begin{equation}
    \ket{E_0} \approx \ket{\Psi_\text{strong}}\,.
\end{equation} 
Exact numerical calculations in small lattice sizes $L\leq8$ confirm that the overlap $|\bra{\Psi_\text{strong}}E_0\rangle|^2\gtrsim 0.9$ with the exact ground state is large. 

The ground state is expected to be a singlet, while the first excited state is anticipated to possess the quantum numbers of the $\sigma$-model particle, namely, a triplet state. Consequently, the operator $d$ must be a component of a vector operator.
Given the form of $\ket{\Psi_\text{strong}}$ in \eq{eq:strong_ground}, we expect the fuzz and head qubits to be treated symmetrically.
We find that the operator\footnote{For $L=4$, one would have
$d_s=(ZI-IZ) II II II - II (ZI-IZ) II II 
+II II (ZI-IZ) II - II II II (ZI-IZ)$.}
\begin{equation}\label{eq:ds}
    d_s = \sum_{l=0}^{L-1}  (-1)^l [Z_{2l+1}\otimes\openone_{2l} - \openone_{2l+1}\otimes Z_{2l}]\,,
\end{equation} connects  $\ket{E_0}$ and $\ket{E_1}$ with significant probability. The initial state is prepared by evolving 
$\ket{\Psi_\text{strong}}$ by the unitary $U=\exp(-i d_s t_\text{prep})$:
\begin{equation}\label{eq:prep_strong}
    \ket{\Psi(0)} = e^{-i d_s t_\text{prep}}\ket{\Psi_\text{strong}}\,,
\end{equation} with a small enough $t_\text{prep}$ such that the contamination from other excited states is minimized.  After that, $\ket{\Psi(0)} $ is evolved by the Hamiltonian  in \eq{eq:fuzzyNew}. It is important to notice that not starting from the exact ground state and/or using an imperfect dipole  operator $d_s \approx d$ does not lead to a change in the measured oscillation frequency. Instead, those imperfections pollute the ideal form \eq{eq:sin} with higher frequencies. If a long time series is available, these frequencies can be separated by Fourier analysis or other methods. Using the current computers where long time series are impossible due to noise/decoherence, is one of the challenges of this method.

In the weak coupling regime $g\alt 0.7$, the ground state can be approximated by $\ket{\Psi_\text{weak}}$ (\eq{eq:weak_ground}). We know that the first excited state lies mostly in the subspace spanned by $\ket{\text{antiferro}} \otimes \ket{\text{any}}$  and the operator connecting it to the first excited state should be a triplet since the ground state is a singlet and the first excited states form a triplet. A simple operator satisfying these conditions is:
\begin{equation}\label{eq:dw}
    d_w = \sum_{l=0}^{L-1} (-1)^{l} Z_{2l+1}\,.
\end{equation}  Again, the initial state is prepared as:
\begin{equation}\label{eq:prep_weak}
    \ket{\Psi(0)} = e^{-i d_w t_\text{prep}}\ket{\Psi_\text{weak}}\,,
\end{equation} for a sufficiently small $t_\text{prep}$, evolved it by the full Hamiltonian in \eq{eq:fuzzyNew} and the expectation value of $d_w$ computed at different times $t$. In the weak coupling limit, the approximate dipole operator $d_w$ has non-negligible transition between the ground and many higher excited states. However, the gap $E_1-E_0$ still dominates the oscillations that can be estimated with appropriate filtering of the oscillations.

\section{Circuits for Quantum Computation }
\label{sec:circuits}

The implementation of the algorithm   above requires state preperation and the translation of the unitary operators into a set of quantum gates. 

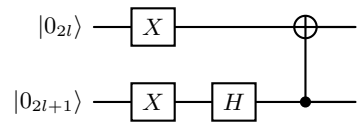
\begin{figure}[tbh]
\begin{center}
\begin{quantikz}
\lstick{$\ket{0_{2l}}$} &\gate{X} & &\targ{}  &\lstick{}  \qw\\
\lstick{$\ket{0_{2l+1}}$} &\gate{X} &\gate{H}  &\ctrl{-1} &\rstick{} \qw
\end{quantikz}
\end{center}
\caption{ Circuit for the singlet Bell state. Wires represent the head ($2l$) and fuzz ($2l+1$) qubits for the site $l$. $H$ is the Hadamard gate.}
\label{circ:singlet}
\end{figure}

We start with the strong coupling limit. The state $\ket{\Psi_\text{strong}}$ is a tensor product of the singlet Bell states comprized of the head and fuzz qubits. The singlet state $(\ket{0_{2l+1}1_{2l}}-\ket{1_{2l+1}0_{2l}})/\sqrt{2}$ can be created by the circuit in Fig.~\ref{circ:singlet}
, where the Hadamard gate $H$
\begin{align}
    H\doteq\frac{1}{\sqrt{2}}\begin{bmatrix}
        1 & 1\\1 &-1
    \end{bmatrix}\,,
\end{align}
is the Hermitian and unitary matrix that diagonalizes $X=HZH$.

The preparation of the initial state $\ket{\Psi(0)}$ and its time evolution to $\ket{\Psi(t)}$  requires the unitary evolution of states involving multiple qubits. This is implemented by applying a sequence of 1- and 2-qubit gates to the multi-qubit gates after decomposing the unitary operations into smaller circuits as we will now show.

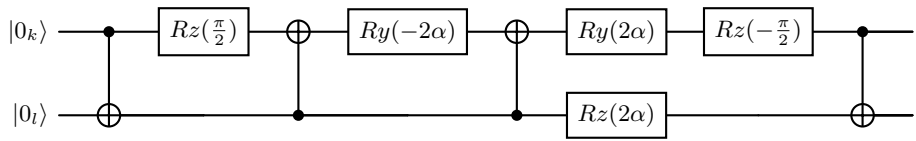
\begin{figure*}
\begin{quantikz}
\lstick{$\ket{0_{k}}$}& \ctrl{1}& \gate{Rz(\frac{\pi}{2})} &\targ{}  &\gate{Ry(-2\alpha)} &\targ{}  &\gate{Ry(2\alpha)} 
& \gate{Rz(-\frac{\pi}{2})} &\ctrl{1}&\rstick{}\qw\\
\lstick{$\ket{0_{l}}$}  & \targ{} &\qw       &\ctrl{-1} &\qw       &\ctrl{-1} & \gate{Rz(2\alpha)} & &\targ{}&\qw \rstick{}\qw
\end{quantikz}
\caption{ Circuit for $\exp[-i\alpha(X_lX_k+Y_lY_k+Z_lZ_k)]$. 
}
\label{circ:RXXYYZZ}
\end{figure*}

A Hermitian operator $M$ can be decomposed into a sum of Pauli strings  $P_i$s such that $M=\sum_i c_i P_i$ with $c_i \in \mathbb{R}$. We have already done this  for the Hamiltonian $H$, and the dipole operators $d_s$ and $d_w$. The  time evolution for a small time step $\Delta t$ is decomposed using the first order Lee-Suzuki-Trotter approximation~\cite{Suzuki:1991}
\begin{align}\label{eq:Trotter}
    e^{i\Delta t  M }= e^{i \Delta t\sum_j c_j P_j}= \prod_{j}e^{i c_j \Delta t P_j}+\mathcal O(\Delta t^2)\,,
\end{align}
whereby the unitary evolution due to the Pauli strings $P_i$s can be applied sequentially. A second order approximation with a smaller error $\mathcal O(\Delta t^3)$ is available, however, it nearly doubles the number of quantum gates which  is not desirable on the noisy machines currently  available. Notice, the Pauli strings in both $d_s$ and $d_w$ commutes, and thus Eqs.~(\ref{eq:prep_strong}), (\ref{eq:prep_weak}) incur no error in their Trotterizations. A longer time evolution is constructed by applying the $\Delta t$ operations multiple times. The constructions of the gates for $\exp(i\Delta t P_j)$ is straightforward since in our calculation the Pauli strings $P_j$s contain at most two Pauli matrices. Implementing $\exp(-i ds \Delta t)$ only requires applying the 1-qubit $R_Z(\theta)=\exp(-i\theta Z/2)$ gate with appropriate $\theta$ on the corresponding qubit. This gate is natively available on both the IonQ-Forte and IBMQ-Fez that we used in our calculations. The weak coupling dipole operator $d_w$ also can be implemented with the $R_Z$ gate.      
We need 2-qubit gates only for the unitary time evolution with the Hamiltonian. 
The gate $R_{ZZ}(\theta)=\exp(-i \theta Z\otimes Z/2)$  is native on the IonQ machines and, 
in combination with the Hadamard and the phase gate
\begin{align}
    S\doteq\begin{bmatrix}
        1 &0\\0 &i
    \end{bmatrix}\,,
\end{align} generate $R_{XX}$ and $R_{YY}$ through the use of the identities $X=HZX$ and $Y=SHZHS^\dagger$.
On the IBMQ-Fez machine, $R_{ZZ}(\theta)$ is available natively as a fractional gate for  $\theta\leq \pi/2$. The $R_{XX}, R_{YY}$ and $R_{ZZ}$ gates can also be implemented as the circuit shown in Fig.~\ref{circ:RXXYYZZ}~\cite{Vatan-Williams:2004} 

The final ingredient we need is measurement of the dipole operator once we created the state $\ket{\Psi(t)}$. The dipole operators $d_s$, $d_w$ are composed of sums of Pauli strings that commute trivially since they are composed of identical Pauli matrices acting on different qubits. Thus, the expectation value of the entire sum of Pauli strings can be done by measuring the qubits in the $Z$ (computational) basis. The expectation value  
\begin{multline}
 \bra{\Psi(t)} Z_l\ket{\Psi(t)}= \bra{q_l(t)}Z_l\ket{q_l(t)}\\
 = P[q_l(t)=0]-P[q_l(t)=1] \,,
\end{multline} where we only need to take the difference in the probabilities $P(q_l)$ in the measurement of the $\ket{0_l}$ and  $\ket{0_l}$ states for all qubits with appropriate phases for $d_s$ and $d_w$ in \eq{eq:ds} and \eq{eq:dw}, respectively.

\section{Results and Discussions }
\label{sec:results}
We separate the discussion of the strong coupling and weak coupling results for clarity. We quote numerical results in lattice units, that is, we scale dimensionful quantities by appropriate powers of lattice spacing $a$ and write them as dimensionless in the rest of the paper.

\subsection{Strong coupling measurements}
\label{subsec:strong}

We performed several calculations on IonQ-Forte and IBMQ-Fez for box sizes from $L=4$ to $L=20$ at coupling $g=1.2$. We prepared the initial state for the time evolution using \eq{eq:strong_ground}, \eq{eq:prep_strong} and 
$t_\text{prep}=0.1$. The time evolution was performed with a Trotter step $\Delta t=0.4$. This choice is a compromise between being able to resolve the dynamics of the model at a scale $\approx a $ and the need to minimize the number of Trotter steps to reduce noise. 
We made 1000 measurements at each time step, enough to make the statistical noise less important than all other approximations.

\begin{table}[tbh]
\centering
\caption{\protect Data from IonQ-Forte, IBMQ-Fez Quantum Processor Unit (QPU) and ideal simulator ($L=12$). The parameters $A$, $\gamma$, $\omega$ were obtained from fits to total time $t=5.2$ at intervals of $\Delta t=0.4$. We use lattice untis.}
\begin{ruledtabular}
\begin{tabular}{ccccc}
 $L$ &  $A$ & $\gamma$ & $\omega$ 
 & QPU
\\ \hline \rule{0pt}{0.9\normalbaselineskip}
\begin{filecontents*}{QPU-Table-2025-12-11.csv}
ll,aa,daa,gg,dgg,omeg,domega,machine
4,-3.97,0.17,0.544,0.031,2.64,0.03,0
6,-5.45,0.23,0.663,0.036,2.73,0.04,0
8,-7.49,0.23,0.534,0.023,2.62,0.02,0
12,-9.83,0.21,0.204,0.009,2.54,0.01,0
8,-6.8,0.23,0.536,0.025,2.48,0.03,1
16,-13.12,0.50,0.975,0.045,2.53,0.05,1
20,-17.47,0.85,1.323,0.072,2.41,0.07,1
12,-9.87,0.17,0.021,0.006,2.38,0.01,2
\end{filecontents*}
\csvreader[head to column names, late after line=\\]{QPU-Table-2025-12-11.csv}{}
{\ \ll 
& \num{\aa\pm\daa} 
& \num{\gg\pm\dgg} 
& \num{\omeg\pm\domega} 
& \ifthenelse{\equal{\machine}{0}}{IonQ}{\ifthenelse{\equal{\machine}{1}}{IBMQ}{Ideal} }
}
\end{tabular}
\end{ruledtabular}
 \label{table:QPU}
\end{table}

\begin{figure}[thb]
\begin{center}
\includegraphics[width=0.48\textwidth,clip=true]{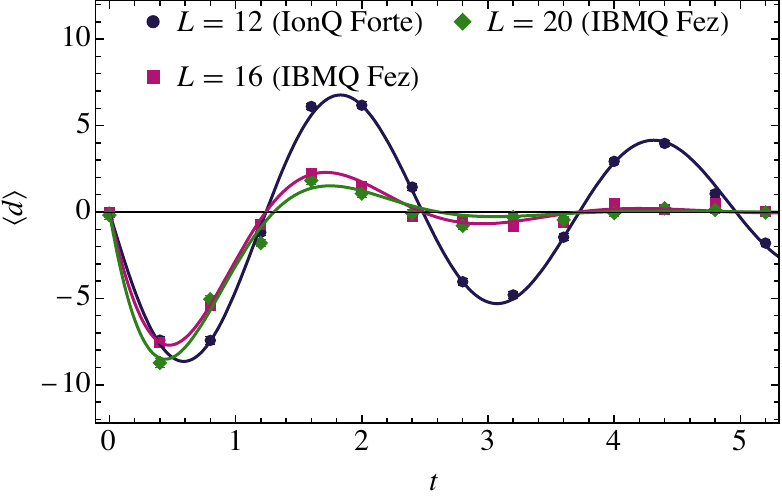}
\end{center}
\caption{\protect  Fits to QPU data. Only the largest 3 lattice box sizes shown. The complete set of fits is in Table~\ref{table:QPU}. Data up to $t=5.2$ were used in the fits. 
}
\label{fig:QPU}
\end{figure}
The resulting measurements at the larger lattices $L=12$, 16 and 20 are shown in Fig.~\ref{fig:QPU}. 
The data exhibits clear oscillatory behavior,  dominated by a single frequency. Furthermore, a pronounced damping of the amplitude is apparent.

There are three reasons why this oscillation does not have  the pure sinusoidal form in \eq{eq:sin}. First, our initial state is not an exact linear combination of ground and first excited states and the contamination of other states with their corresponding frequencies is to be expected. Second, the time evolution contains Trotterization errors due to the finite value of $\Delta t$. Finally, current devices are not error corrected and suffer from noise and decoherence errors.
Simulations of the same calculation in classical computers and $L\leq 10$, which include Trotter errors and the contamination of excited states, show no sign of damping. Therefore, we attribute the damping shown in Fig.~\ref{fig:QPU} to noise and decoherence of the quantum devices we used.

 A fit of the ideal simulation results to a damped sinusoidal form 
\begin{equation}\label{eq:fit_damped}
   \langle d\rangle = A e^{-\gamma t}\sin \left( \omega t \right) \,,
\end{equation} gives
the values in \tab{table:QPU}. The errors shown in \tab{table:QPU} reflect only the quality of the fit to \eq{eq:fit_damped} and the uncertainty due to the finite number of shots. 
The quality of these fits are exemplified by the solid lines  in Fig.~\ref{fig:QPU}.
The goodness of the fit is another indication that, at least for the parameters considered, there is very little contamination of excited states outside the subspace spanned by $\ket{E_0}, \ket{E_1}$. This is true despite the fact that neither $\ket{\Psi_\text{strong}}$ equals the exact ground state $\ket{E_0}$ nor that $d\ket{E_0}$ equals the exact excited state $\ket{E_1}$. 
Simulations in classical computers show that the contamination from excited states would be very apparent if larger
 values of $t_\text{prep}$ were used and this guided our choice of $t_\text{prep}$. 

The fitting function in \eq{eq:fit_damped} fits the data fairly well but is arbitrary. Therefore, 
 the identification of $\omega$ appearing in \eq{eq:fit_damped} with the mass gap $m$ introduces an uncontrolled error arising from this arbitrariness. The dependence of the mass gap $m$ with the volume is known to be suppressed by $\sim e^{-mL}$ in the $m L \rightarrow\infty$ limit \cite{luscher-1}. The values of $\omega$ shown in \tab{table:QPU} are clearly in this regime and, therefore, we do not expect  a noticeable volume dependence on the value of the gap. The slight discrepancy between the values of $\omega$ in \tab{table:QPU} should then be ascribed to the uncontrolled errors due to the arbitrary form in \eq{eq:fit_damped} and serve as a measure of this uncertainty. The average of $\omega$ over the different values of $L$ from the Quantum Processor Unit runs results in
 $\omega= \num{2.56\pm0.02}$, where the error is given by the spread in the values in \tab{table:QPU}. Attempts to reduce and quantify the Trotterization errors by performing calculations at smaller Trotter steps $\Delta t$ turned out to be impossible due to the significant increase of noise/decoherence errors as the number of Trotter steps increases.

 If we identify $\omega$ in \eq{eq:sin} with the gap $m=E_1-E_0$
we find a discrepancy from the correct value $m_{exact} \approx 2.34$ obtained from direct diagonalization in lattices $L\leq12$ (the finite volume effect beyond $L=10$ are negligible) of about $9\%$. As we will see below, a noiseless calculation, for which \eq{eq:fit_damped} is adequate, eliminates this discrepancy.

\subsection{Weak coupling analysis}
\label{subsec:weak}

\begin{figure*}
\begin{center}
\includegraphics[height=0.3\textwidth,clip=true]{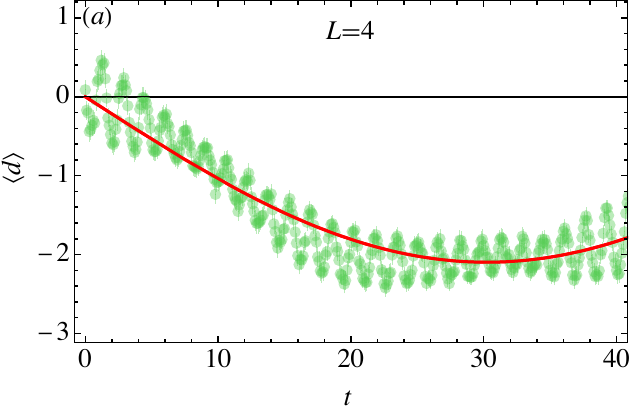} \hfill
\includegraphics[height=0.3\textwidth,clip=true]{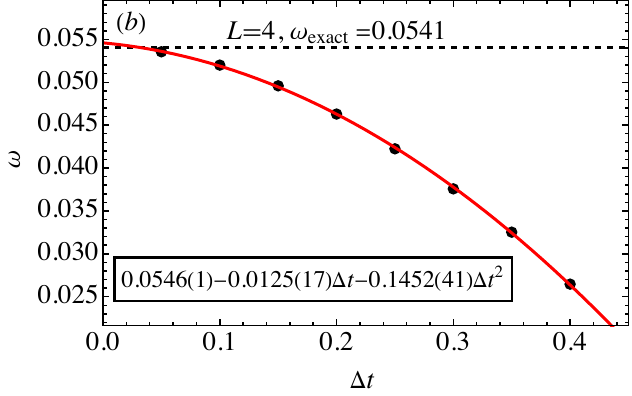}\\\vspace{0.2in}
\includegraphics[height=0.3\textwidth,clip=true]{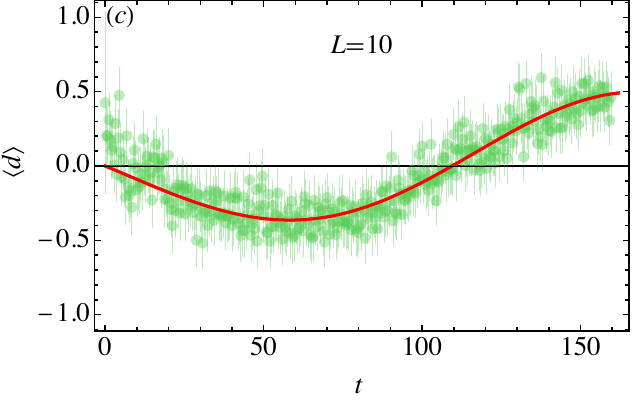} \hfill
\includegraphics[height=0.3\textwidth,clip=true]{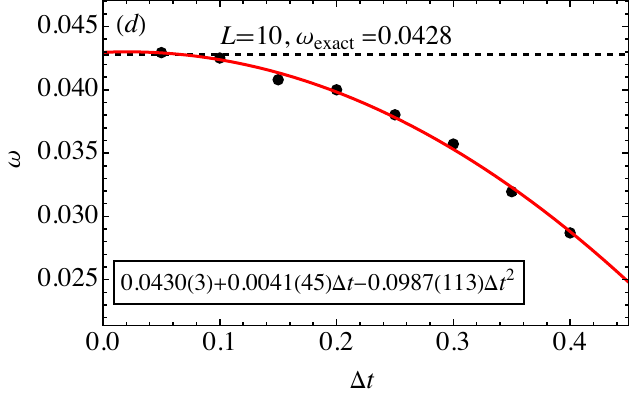}
\end{center}
\caption{\protect Top left panel: Large volume $L=4$ ideal simulation data (green data points) for $\langle d\rangle$ vs time $t$ at weak coupling $g=0.6$ with $\lambda=0.5$ in \eq{eq:extrapolate}, $t_\text{prep}=0.1$, and Trotter steps $\Delta t =0.1$. 
The solid (red) curve is a fit of the form in \eq{eq:fit_damped} whose fit parameters are provided in the text. 
Top right panel: Extrapolation of the gap $\omega$ to $\Delta t\rightarrow 0$  for $g=0.6$, $L=4$. The (black) data points are the gaps extracted from ideal simulations at $\Delta t$ values as indicated.  The solid (red) curve is a polynomial fit whose coefficients are indicated in the boxed insert. The extracted value $\omega=\num{0.05461\pm0.0001}$ is to be compared with the exact numerical result $\omega_\text{exact}=0.0541$. We have indicated only the errors from the $\chi^2$ fit. As discussed in the text, a cubic $\Delta t^3$ term in the fit brings the extracted  $\omega$ value in better agreement with the exact result. 
Bottom left panel: Large volume $L=10$ ideal simulation data (green data points) for $\langle d\rangle$ at weak coupling $g=0.6$ with $\lambda=0.1$ in \eq{eq:extrapolate}, $t_\text{prep}=0.01$, and Trotter steps $\Delta t =0.4$.  Rest of the notation is the same as the top left panel. 
The solid (red) curve  fit parameters are provided in the text. 
Bottom right panel: Extrapolation of the gap $\omega$ to $\Delta t\rightarrow 0$  for $g=0.6$, $L=10$.  Rest of the notation is the same as the top
right panel. 
 The solid (red) curve is a polynomial fit with coefficients as indicated in the boxed insert. The extracted value $\omega=\num{0.0430\pm0.0003}$ is in agreement with the exact numerical result $\omega_\text{exact}=0.0428$. We have indicated only the errors from the $\chi^2$ fit. 
}
\label{fig:weakDtExtrapolation}
\end{figure*}

The weak coupling limit, which includes the scaling region near the continuum limit, presents a different kind of challenge. Near the continuum limit the gap is very small (in lattice units) and the oscillation very slow. Therefore, one needs to compute $\langle d \rangle$ to much larger times. However, in order to capture the high momentum modes correctly, the Trotter step size $\Delta t$ has to remain small. Consequently, the number of Trotter steps required is much larger. In an error corrected machine this would lead to a moderate cost increase,  linear on the inverse gap. In a noisy machine, however, a large number of Trotter steps means a large number of gates and the dilution of the signal by noise/decoherence before any meaningful results can be extracted. Alternatively to Trotter, other simulation algorithms, such as the quantum signal processing (QSP)~\cite{PhysRevLett.118.010501}, scale better but are prohibitive on current noisy hardware.
For this reason, we demonstrate the feasibility of the general algorithm in an ideal (noiseless) simulation in Fig.~\ref{fig:weakDtExtrapolation} at $g=0.6$ 
using the Trotter decomposition for simplicity.

The weak coupling calculation  requires a few additional considerations. As the coupling gets smaller and the correlation lengths larger, 
the finite volume corrections become more noticeable: whereas at the strong coupling $g=1.2$, the gap varies from $\omega=2.3368$ at $L=4$ to $\omega=2.3370$ at $L=10$; at weak coupling $g=0.6$, we find the gap varies from $\omega=0.054098$ at $L=4$ to $\omega=0.042759$ at $L=10$. Usually, larger volume simulations would be desirable for $L\rightarrow\infty$ extrapolations using L{\"u}scher method. The initial state $|\Psi(0)\rangle$ preparation from $|\Psi_\text{weak}\rangle$ also requires some deliberation. We construct a dipole operator
\begin{multline}\label{eq:extrapolate}
    d(\lambda) = \lambda d_s+(1-\lambda) d_w\\
    =\sum_{l=0}^{L-1}(-1)^l
[Z_{2l+1}\mathbbm{1}_{2l}  -\lambda \mathbbm{1}_{2l+1} Z_{2l}]\,,
\end{multline}
from a simple linear extrapolation between the strong and weak coupling limits. Further, we take a smaller $t_\text{prep}$ when $L$ is larger to reduce the contamination from the higher excited states to which $d_w$ couples.

We perfom a small $L=4$ and a large $L=10$ volume calculation at weak coupling $g=0.6$ to illustrate what has been discussed above, see Fig.~\ref{fig:weakDtExtrapolation}. At $L=4$, we prepare the initial state $\ket{\Psi(0)}$ using $t_\text{prep}=0.1$ and $\lambda=0.5$  with Trotterization steps $\Delta t=\numrange{0.05}{0.4}$ in intervals of 0.05.
The contamination from higher excited states in Fig.~\ref{fig:weakDtExtrapolation} $(a)$ for the $\Delta t=0.1$ result is visible as fast oscillations on top of the dominant feature due to the low frequency oscillations from the vanishing energy gap $E_1-E_0$.   We fit the form in \eq{eq:fit_damped} and 
find 
$A=\num{-2.10\pm0.02}$, $\omega=\num{0.0520\pm0.00005}$ with a vanishing $\gamma=\num{-9.6\pm8.2}\times 10^{-5}$ 
damping exponent, indicating  the sinusoidal nature of the signal.

The larger volume $L=10$ calculation in Fig.~\ref{fig:weakDtExtrapolation} was performed with an initial state 
$\ket{\Psi(0)}$  prepared with a much shorter $t_\text{prep}=0.01$ and smaller $\lambda=0.1$  to eliminate contamination from higher excited state. As the volume increases, the energy levels shift downward at different rates, decreasing the relative gaps between states in the process. A shorter  $t_\text{prep}$ and smaller  $\lambda$ becomes necessary to obtain a dominant sinusoidal form. A consequence of shorter  $t_\text{prep}$ is smaller overlap between the ground and excited states though one can still obtain the desired signal. We find a fit to simulation data using \eq{eq:fit_damped}  gives $A=\num{-0.309\pm0.014}$, $\gamma=\num{-0.0029\pm 0.0004}$, $\omega=\num{0.0287\pm0.0002}$ at $\Delta t=0.4$. In Fig.~\ref{fig:weakDtExtrapolation} $(c)$, we show results for $\Delta t=0.4$ instead of $\Delta t =0.1$  at $L=10$ for visualization as the latter would have four times as many data points in the same time interval. We also mention that both the $L=4$ and $L=10$ data were fit over the same time interval of $t=160$.  Fig.~\ref{fig:weakDtExtrapolation} $(a)$ is zoomed in  over a shorter time interval to highlight the fast oscillations.

In Fig.~\ref{fig:weakDtExtrapolation}, right panels, we show extrapolations of ideal simulation data to $\Delta t\rightarrow 0$. 
At $L=4$, a quadtratic polynomial fit gives an extrapolated gap 
$\omega=\num{0.0546\pm 0.0001}$  that changes to $\omega=\num{0.0542\pm 0.0001}$ 
when a cubic $\Delta t^2$ term is added. This is to be compared to $\omega_\text{exact}=0.0541$. 
A similar fit at $L=10$ gives an extrapolated gap $\omega=\num{0.0430\pm 0.0003}$ in good agreement with the exact numerical result $\omega_\text{exact}=0.0428$. 
Adding a cubic term changes the fitted value slightly to $\omega=\num{0.0435\pm 0.0006}$. 
 All the quoted errors here are from the fits only. 

\begin{figure}[thb]
\begin{center}
\includegraphics[width=0.48\textwidth,clip=true]{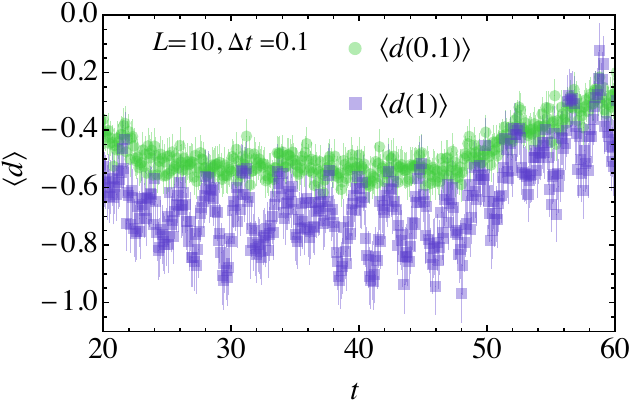}
\end{center}
\caption{\protect Measurement of dipole operators $d(\lambda=0.1)$ and $d(\lambda=1)$ using the same state $|\Psi(t)\rangle$ prepared from $|\Psi(0)\rangle=\exp[-i t_\text{prep}\, d(0.1)]|\Psi_\text{weak}\rangle$ with $t_\text{prep}=0.1$ and $\Delta t =0.1$ at $g=0.6$ and $L=10$.}
\label{fig:weakHighFreq}
\end{figure}
Before concluding, we make the following observation. The algorithm we presented (for both strong and weak coupling) involves several distinct steps. First, we prepare the initial state $|\Psi(0)\rangle$ with an unitary evolution with a dipole operator for time $t_\text{prep}$. The form of the dipole operator whether $d_s$, $d_w$ or some linear combination of these allows error free Trotterization since each Pauli strings in the dipole operators mutually commute. However, as discussed earlier, the choice of $t_\text{prep}$ depends on the size of the coupling $g$, with a smaller $t_\text{prep}$ at weaker couplings to reduce contamination from higher excited states. The second step involves unitary evolution with the system Hamiltonian which incurs Trotterization errors. Thus, one has to balance the Trotter steps $\Delta t$ with the circuit depth especially at weak couplings where a longer total time evolution becomes necessary for the vanishing gap. The final step involves measurement of the dipole operator. The only necessary condition to produce the oscillations used in the gap calculation is that we pick an observable/operator that does not commute with the Hamiltonian (and that the initial state is a linear superposition of energy eigen states). In particular, it means we do not have to measure the same dipole operator that was used in the state preparation. In Fig.~\ref{fig:weakHighFreq} we show the expectation values of two different dipole operator using the same exact initial state preparation. We emphasize that this does not require separate quantum simulation. The same simulation can be used for measurement of multiple observables, at post processing. For example, in our algorithm, we produced the initial state $|\Psi(0)\rangle$ at $g=0.6$, $L=10$ with $d(\lambda=0.1)$ and $t_\text{prep}=0.1$ and evolved the system with $\Delta t=0.1$ and recording the state $|\Psi(t)\rangle$ at later time $t$ (in computational basis). We used 10,000 shots to improve precision in the ideal simulation. Then we can measure multiple observables $\langle A\rangle$, $\langle B\rangle$, etc. with the same $|\Psi(t)\rangle$. In Fig.~\ref{fig:weakHighFreq}, the green data points corresponds to measurement of $d(\lambda=0.1)$ where the contamination from the higher excited states are washed out. A fit of the form in Eq.~(\ref{eq:fit_damped}) gives $\omega=\num{0.04296\pm0.00003}$. 
The purple data points are from measurement of $d(\lambda=1)$ that couples strongly to some of the higher excited state resulting in the higher frequency oscillations on top of the dominant low frequency oscillation. A similar fit to Eq.~(\ref{eq:fit_damped}) gives $\omega=\num{0.04268\pm0.00005}$ in agreement with the earlier green data points. This ability to measure different observables from the same state preparation and unitary time evolution is not a shortcoming but an advantage. It can be used to highlight different aspect of the physical system.

\section{Conclusion}
The results presented here demonstrate a viable method for calculating the mass spectrum in a field theory using quantum computers. By constructing an operator that couples the 
lowest energy states with two different quantum numbers and
 tracking the time evolution of its expectation value, we can extract the mass gap from the oscillation frequency. Using exact diagonalization on small lattices, we verified this approach and then extended it to larger systems beyond classical computational reach.

Our method was shown to work for lattice sizes up to $L = 20$, corresponding to a Hilbert space of dimension $2^{40}$, well beyond the capacity of direct diagonalization in a classical computater. Although the signal becomes increasingly damped with system size, as long as the gap remains sufficiently large, the oscillation frequency can still be accurately fitted. 
In the continuum/scaling region at weak coupling,
 the oscillations become slower and the signal decays before a complete oscillation in $\langle d \rangle$ can be observed. In this regime, better, error corrected computers would be required to apply our method.

Overall, this work establishes a proof of concept for probing excitation spectra and real-time dynamics in lattice field theories using quantum devices. Demonstrating simulations of systems with Hilbert spaces far exceeding classical limits represents a significant step toward leveraging quantum hardware for nonperturbative field theory calculations.

\acknowledgments
This work was partially supported 
 by U.S. DOE Grant No.  DE-SC0024286 and   NSF Grant PHY-2209184.
 We acknowledge the National Quantum Laboratory (QLab) at the University of Maryland for providing access to the IonQ Forte system for this work. Part of the work was completed by GR at the Maryland Center for Fundamental Physics, Univ. of Maryland, College Park during his sabbatical. We acknowledge the use of IBM Quantum services for this work. In
this paper we used \verb|ibmq_fez|.

%

\end{document}